\documentclass[twocolumn]{aastex63}

\let\bfseries\mdseries
\received{X N, 2019}
\revised{Y N, 2019}
\accepted{\today}
\submitjournal{ApJ}

\shorttitle{A HaloSat Analysis of the Cygnus Superbubble}
\shortauthors{Bluem et al.}
\graphicspath{{./}{figures/}}

\begin{document}

\title{A HaloSat Analysis of the Cygnus Superbubble}

\email{jesse-bluem@uiowa.edu}

\author{Jesse Bluem}
\affiliation{University of Iowa Department of Physics and Astronomy, Van Allen Hall, 30 N. Dubuque St., Iowa City, IA 52242, USA}

\author{Philip Kaaret}
\affiliation{University of Iowa Department of Physics and Astronomy, Van Allen Hall, 30 N. Dubuque St., Iowa City, IA 52242, USA}

\author{William Fuelberth}
\affiliation{University of Iowa Department of Physics and Astronomy, Van Allen Hall, 30 N. Dubuque St., Iowa City, IA 52242, USA}

\author{Anna Zajczyk}
\affiliation{University of Iowa Department of Physics and Astronomy, Van Allen Hall, 30 N. Dubuque St., Iowa City, IA 52242, USA}
\affiliation{NASA Goddard Space Flight Center, Greenbelt, MD 20771, USA }
\affiliation{Center for Space Sciences and Technology, University of Maryland, Baltimore County, 1000 Hilltop Circle, Baltimore, MD 21250, USA} 

\author{Daniel M. LaRocca}
\affiliation{University of Iowa Department of Physics and Astronomy, Van Allen Hall, 30 N. Dubuque St., Iowa City, IA 52242, USA}

\author{R. Ringuette}
\affiliation{University of Iowa Department of Physics and Astronomy, Van Allen Hall, 30 N. Dubuque St., Iowa City, IA 52242, USA}

\author{Keith M. Jahoda}
\affiliation{NASA Goddard Space Flight Center, Greenbelt, MD 20771, USA }

\author{K. D. Kuntz}
\affiliation{NASA Goddard Space Flight Center, Greenbelt, MD 20771, USA }
\affiliation{The Henry A. Rowland Department of Physics and Astronomy, Johns Hopkins University, 3701 San Martin Dr., Baltimore, MD 21218, USA}



\begin{abstract}
The Cygnus Superbubble (CSB) is a region of soft X-ray emission approximately 13 degrees wide in the direction of the local spiral arm. Such a large region might be the result of strong stellar winds and supernovae from nearby stellar nurseries, or it could be the result of a single event - a hypernova. HaloSat observed 4 non-overlapping 10 degree diameter fields in the CSB region over the 0.4-7 keV band. The CSB absorption and temperature was found to be consistent over all 4 fields, with a weighted average of $\rm 6.1 \times 10^{21}$ $\rm cm^{-2}$ and $0.190$ keV, respectively. These observations suggest that the CSB is a cohesive object with a singular origin. The total thermal energy for the CSB is estimated at $\rm 4 \times 10^{52}$ erg, based upon a shell-like physical model of the CSB. \textbf{Absorption and distance estimates to Cyg OB associations are examined. The CSB absorption is found to be most consistent with the absorption seen in Cyg OB1, implying that the CSB lies at a similar distance of 1.1-1.4 kpc.}

\end{abstract}



\section{Introduction} \label{sec:intro}

In 1980, an extended soft X-ray structure was discovered by HEAO 1 near the plane of the galaxy, in the direction the constellation Cygnus \Citep{Cash1980}. This region of soft X-rays was found to be related to previous IR, optical, and radio structures observed in the same region, and is now commonly referred to as the Cygnus Superbubble (CSB). \citet{Cash1980} found the X-ray emission was spread over 13 degrees of the sky, corresponding to a diameter of 450 pc at roughly 2 kpc distance, as estimated by absorption measurements. The CSB appears to resemble a horseshoe at first glance, but this is an artifact induced by the intervening Cygnus Rift (or, if the reader prefers, the Northern Coalsack or Great Rift of Cygnus), a large dust cloud that obscures the central region of the CSB. Near the CSB are 9 OB associations, including the notable Cygnus OB2 association. Cygnus OB2 contains more than 100 O spectral class stars, making it the largest such assemblage of O stars and the highest mass of any young stellar association yet detected in our galaxy \citep{Knodlseder2000}.

When we look towards the CSB, we are looking down the length of the local spiral arm. Separate objects along this line of sight are superposed on this region of the sky, making it difficult to determine if observed structures are discrete objects or multiple superposed objects. This effect, combined with conflicting measurements of the distance to regions of the bubble, has made understanding the precise nature of the CSB difficult.

One way to study distance is to look at absorption, typically \textbf{parameterized by} the total hydrogen column density. The farther away an object is, the more intervening Galactic material absorbs light radiated by the object. In the case of the CSB, conflicting measurements of $\rm N_H$ have resulted in evidence supporting both composite and discrete origins for the observed structure. \citet{Uyaniker2001} found differing $\rm N_H$ values for different regions of the CSB, implying that the CSB is a composite object dependent upon our particular line of sight along the spiral arm. However, more recently \citet{Kimura2013} has found similar $\rm N_H$ values across the CSB - inferring that the CSB is a unified structure and not a line of sight composite.

If the CSB is a unified structure, then explaining the large size of the region becomes problematic. \Citet{Cash1980} estimated the total thermal energy of the CSB as exceeding $\rm 6 \times 10^{51}$ erg, for a distance of 2 kpc, and favor an interpretation of a series of 30-100 supernovae rather than a single event. If the CSB did originate from a singular event, it would have to be a very rare sort of supernova, referred to as a hypernova \citep{Paczynski1998}. Some observational evidence of these hypernovae exist. SN1998bw is a supernova with an estimated $\rm 2-5 \times 10^{52}$ erg initial kinetic energy, putting it an order of magnitude above other supernova, and may have stemmed from a progenitor star with a mass of 40 solar masses \citep{Iwamoto1998}. This observed energy is similar to the energy observed in the CSB.

Another possibility is a combination of multiple supernovae and/or stellar winds from the OB associations in the area. Cygnus OB2 in particular could be involved, being one of the most impressive star formation regions in the galaxy. An analysis of the Cyg OB2 association found significant mass loss rates for some of the most notable members of the association, and that such a strong wind could conceivably generate a bubble on the scale of the CSB in 2 million years \citep{Abbott1981}. \citet{Cash1980} suggests that the appearance of a wind-blown bubble could be achieved by 30-100 supernova over the preceding 3-10 million years, meaning that the current most massive Cyg OB2 association stars must be from a younger generation than the progenitors of these novae. However, given that Cygnus OB2 is offset from the center of the fairly circular CSB, it is less likely to be the direct source of the bubble, be it from winds or supernovae. This is not an issue for a hypernova - the offset could be explained by a runaway star ejected from Cyg OB2 \citep{Kimura2013}.

Observations of the CSB were taken by HaloSat in October 2018, September 2019, and October 2019. Halosat is specifically suited for studying diffuse emission, with an observing program that emphasizes reducing foreground contamination. The HaloSat energy range goes down to 0.4 keV, allowing for good measurements of absorption. Section \ref{sec:obs} describes the details of these observations. Section \ref{sec:ana} covers the spectral analysis. Section \ref{sec:dis} details estimating a distance to the CSB by using fitted absorption and previous literature on the Cyg OB associations. Section \ref{sec:res} describes the spectral results and estimates global parameters of the CSB using the estimated distance.

\section{Observations} \label{sec:obs}

\begin{figure*}[htb!]
\centering
\includegraphics[width=1\textwidth]{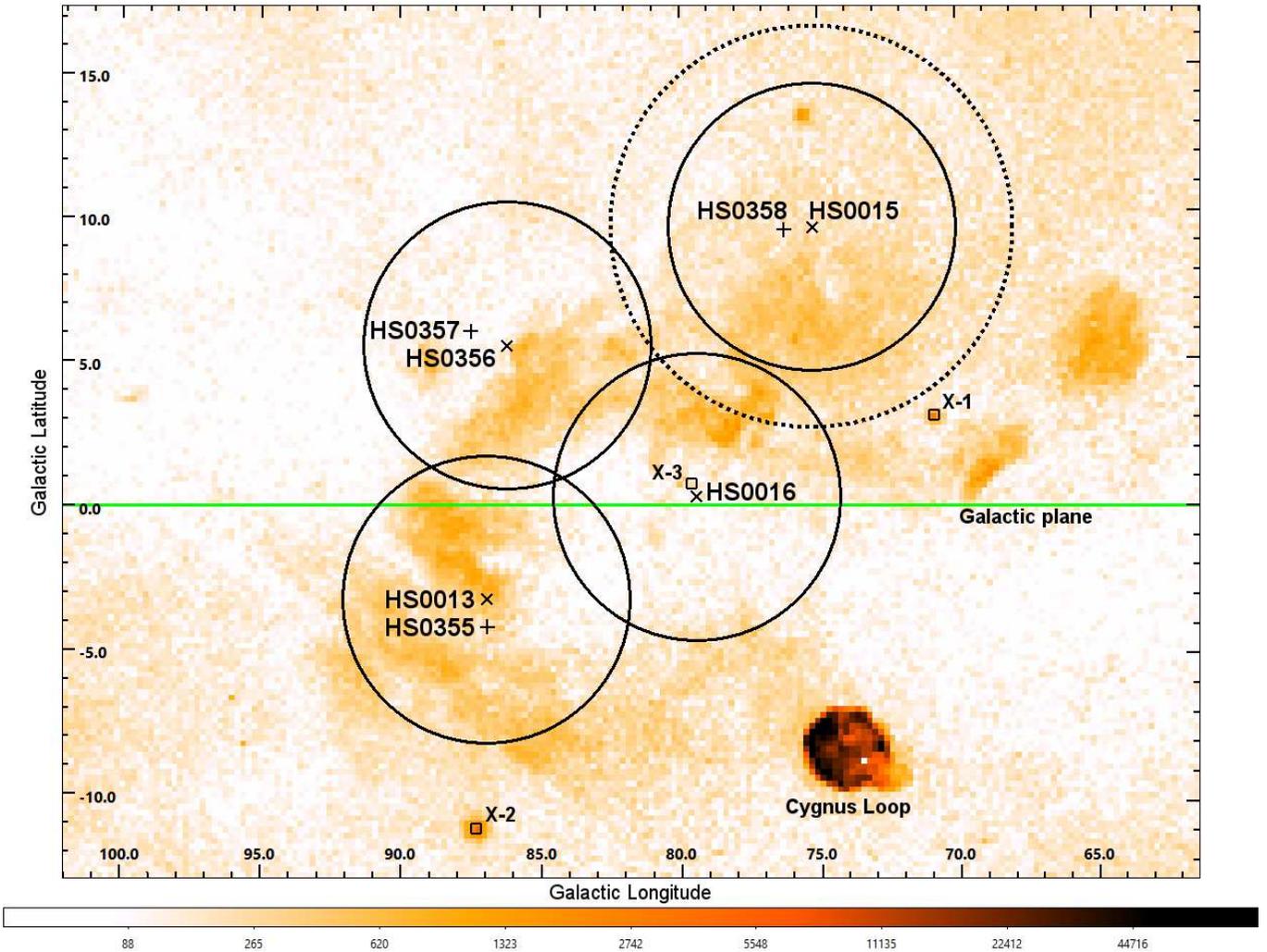}
\caption{ROSAT 3/4 keV (R4+R5) map \citep{Snowden1997} showing the HaloSat CSB fields. Primary targets are marked with an x and offset targets are marked with a cross. Cygnus X-1, X-2, and X-3 are marked with boxes. The plane of the Milky Way is the line through the middle of the figure. For each field, the solid circle is the HaloSat 10 degree diameter full-response field of view. Only the fields of view for the primary targets are shown. The dashed circle is an example of where the HaloSat response drops to zero (14 degree diameter). \textbf{The color bar is in units of $\rm 10^{-6}$ ct $\rm s^{-1}$ $\rm arcmin^{-2}$.}}
\end{figure*}

HaloSat is a NASA funded CubeSat with instrumentation developed at the University of Iowa and was deployed from the International Space Station on 2018 July 13 \citep{Kaaret2019, LaRocca2020}. Using three silicon drift detectors, HaloSat specializes in observing diffuse X-ray emission, \textbf{including} extended sources. The average energy resolution of these detectors is 85 eV at 676 eV (fluorine K$\rm \alpha$) and 137 eV at 5.9 keV (manganese K$\rm \alpha$) \citep{Kaaret2019}. HaloSat has full response over a 10 degree diameter field of view, which then falls off linearly to zero at 14 degrees. 

The CSB was covered with four initial fields \textbf{(HS0013, HS0014, HS0015, HS0016),} selected with the intent of covering as much of the CSB as possible. The pointing coordinates for the fields were chosen to avoid the strong X-ray sources Cygnus X-1 and X-2, as well as to avoid the nearby Cygnus Loop, which was itself a separate target observed by HaloSat. Early observations were taken with a misalignment between the spacecraft's boresight and the intended observation target, resulting in a small offset of 1 degree. These early offset fields (HS0355, HS0357, HS0358) were given their own HaloSat IDs and analyzed separately from the later observations. HS0014's pointing was moved before any data was taken, becoming HS0356 and resulting in no observations being taken under the HS0014 ID. The HaloSat fields can be seen overlaid on a ROSAT 3/4 keV (R4+R5) map \citep{Snowden1997} in Figure 1.

\begin{deluxetable*}{cccccccc}
\tablenum{1}
\tablecaption{Observation Parameters \label{tab:MP}}
\tablewidth{0pt}
\tablehead{
\colhead{HS ID} & \colhead{RA}  & \colhead{dec} & \colhead{nadir} & \colhead{solar angle} & \colhead{detector} & \colhead{exposure} & \colhead{hard rate}\\
\colhead{} & \colhead{(deg)} & \colhead{(deg)} & \colhead{(deg)} & \colhead{(deg)} && \colhead{(s)} & \colhead{(c $\rm s^{-1}$)} 
}
\startdata
HS0013 & $318.49$ & $+43.88$ & $145$ & $124$ & 14 & 12224 & 0.076\\
 &  &  &  & & 38 & 12480 & 0.066\\
 &  &  &  & & 54 & 12928 & 0.070\\
HS0355 & $319.43$ & $+43.18$ & $116$ & $116$ & 14 & 16320 & 0.053\\
 &  &  &  & & 38 & 15936 & 0.048\\
 &  &  &  & & 54 & 15680 & 0.050\\
\hline
HS0356 & $307.97$ & $+48.99$ & 144, 174 & 116, 112 & 14 & 13888 & 0.097\\
 &  &  &  & & 38 & 10310 & 0.098\\
 &  &  &  & & 54 & 17216 & 0.090\\
HS0357 & $308.48$ & $+50.29$ & $118$ & $108$ & 14 & 19264 & 0.058\\
 &  &  &  & & 38 & 19136 & 0.052\\
 &  &  &  & & 54 & 19200 & 0.054\\
\hline
HS0015 & $294.96$ & $+42.04$ & $131$ & $114$ & 14 & 6656 & 0.082\\
 &  &  &  & & 38 & 8256 & 0.085\\
 &  &  &  & & 54 & 8128 & 0.088\\
HS0358 & $295.67$ & $+42.87$ & $144$ & $105$ & 14 & 11584 & 0.065\\
 &  &  &  & & 38 & 11328 & 0.067\\
 &  &  &  & & 54 & 11456 & 0.064\\
\hline
HS0016 & $308.43$ & $+40.57$ & 151 & 117 & 14 & 30144 & 0.086\\
 &  &  &  & & 38 & 31616 & 0.083\\
 &  &  &  & & 54 & 32384 & 0.084\\
\enddata
\tablecomments{Earlier offset fields are paired with their later counterpart fields. Column 1 is the HaloSat field ID, Column 2 and 3 are right ascension and declination. Column 4 and 5 are average nadir angle and average solar angle for the duration of the observation. Column 6, 7, and 8 are specific to each detector for each observation. Column 6 is the detector unit ID, column 7 is exposure time for each detector after cuts, and column 8 is the hard count rate in the detector.}
\end{deluxetable*}

Cuts were performed based on a 0.16 ct $\rm s^{-1}$ count rate in the HaloSat ``hard" band (3-7 keV) and a 0.75 ct $\rm s^{-1}$ count rate in the HaloSat ``very large event" band (7+ keV). Some observations with high particle-induced backgrounds (HS0359 and HS0360) were removed due to significant data loss from the cuts used. Table 1 includes the observation parameters for each field, including right ascension and declination for the field pointing, nadir angle, and anti-Sun angles at the time or times of the observations, and the exposure time and hard count rate for each detector after cuts were applied.

\section{Analysis} \label{sec:ana}

\begin{figure*}
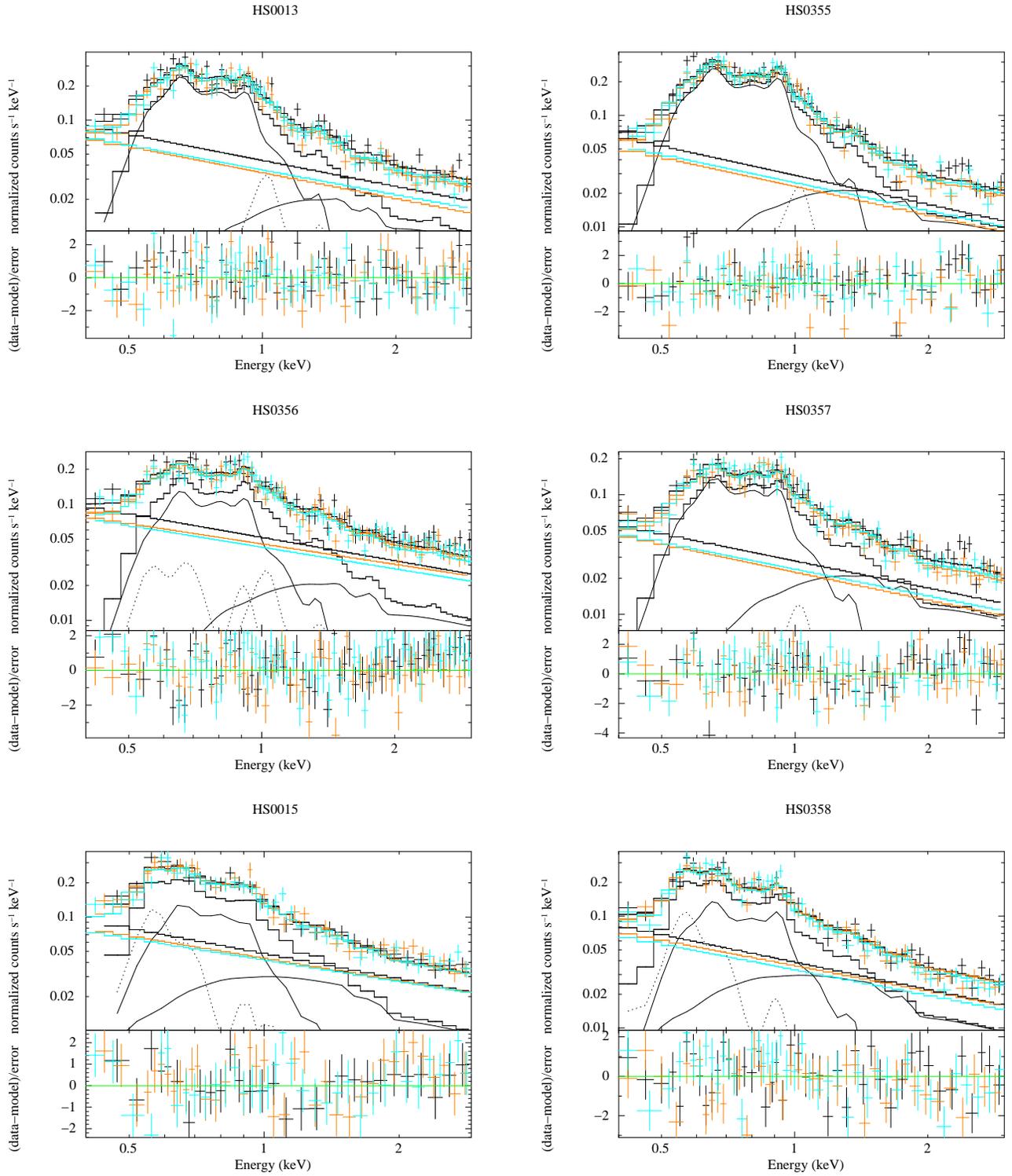

\centering
\gridline{\includegraphics[width=0.35\textwidth,angle=270]{13replot.eps}
          \includegraphics[width=0.35\textwidth,angle=270]{355fitfix.eps}
          }
\gridline{\includegraphics[width=0.35\textwidth,angle=270]{356replot.eps}
          \includegraphics[width=0.35\textwidth,angle=270]{357replot.eps}
          }
\gridline{\includegraphics[width=0.35\textwidth,angle=270]{15replot.eps}
          \includegraphics[width=0.35\textwidth,angle=270]{358replot.eps}
          }
\caption{Spectra for the HaloSat CSB fields. Overlapping fields are adjacent to each other. Each field is labeled with its HaloSat ID at the top of the spectrum. Each detector is represented with a different color. Detector 14 is black, 38 is orange, and 54 is teal. The {\tt apec} is the dominant component in each field, marked with a thick solid line. The NEI background components are marked with dashed lines. All other background components are marked with thin solid lines.
\label{fig:pyramid}}
\end{figure*}

The CSB spectra were analyzed with XSPEC version 12.10.1f \citep{Arnaud1996}. For each field, the CSB is modeled as collisionally-ionized diffuse gas using an absorbed {\tt apec} model \citep{Smith2001}. Also included are multiple background/foreground components. These include detector specific power-laws for the particle backgrounds, fixed models for the absorbed cosmic X-ray background (CXB) and unabsorbed local hot bubble (LHB), and absorbed non-equilibrium ionization (NEI) models for Galactic ridge emission. Field HS0016 includes absorbed model components for Cygnus X-3 (Cyg X-3) found from fitting MAXI data over the same time interval. Due to the CSB being in the Galactic plane, and the high column densities therein, the Galactic halo is not included in the modeled background. All absorptions used the {\tt tbabs} model and the associated Wilms abundances \citep{Wilms2000}. All model components are folded through the HaloSat response matrix, except for the instrumental particle backgrounds \citep{Kaaret2019,Zajczyk2020}.

\begin{figure*}[htb!]
\centering
\includegraphics[width=0.5\textwidth,angle=270]{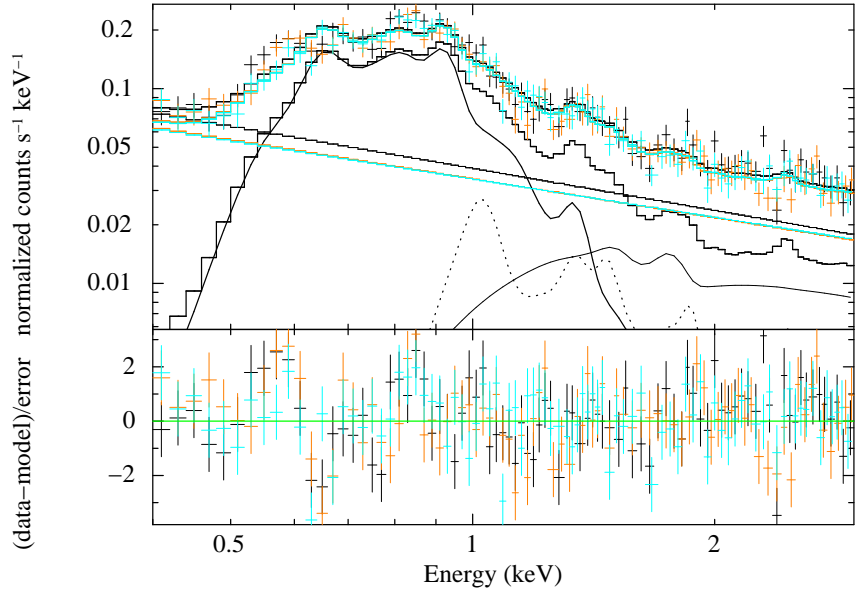}
\caption{Spectrum for the HaloSat CSB field HS0016. Each detector is represented with a different color. The {\tt apec} is the dominant component in the field, marked with a thick solid line. The NEI background components are marked with dashed lines. All other background components are marked with thin solid lines.}
\end{figure*}

The {\tt apec} model used to represent the CSB is fit with freed absorption, temperature, and normalization. The abundances are fixed to 0.26, as found for the CSB in \citet{Kimura2013}. The spectra for each HaloSat field can be seen in Figures 2 and 3 and the fitted model parameters can be found in Table 2. 

The particle backgrounds are power-laws (XSPEC model {\tt powerlaw}) specific to each detector, with fixed photon indices, unique to each detector, calculated using an empirical relation for HaloSat's detectors based on an analysis of the southern halo (Galactic latitude $\rm <$ -30) \citep{Kaaret2020}. The power-law normalizations are fit between 3 and 6.9 keV (some fields do not have enough data for the bins used to extend to 7 keV, so a limit of 6.9 keV is used for all fields) and then fixed before any other model parameters are fit. All other model parameters are linked together between the three detectors. The exceptions to this fitting procedure are fields HS0357 and HS0016. The calculated power-law photon indices for HS0357 \textbf{do not match the slope of the data}, so the photon indices are left free and are fit over the full data range of 0.4 to 6.9 keV while simultaneously fitting the other model parameters. The HS0016 power-law normalizations are also fit over the full range, simultaneously with the other model parameters. Fitting the power-law at the high energies alone is problematic because the Cyg X-3 components affect the full energy range. \textbf{For all fields,} all particle background power-law components are fixed, and the energy range is narrowed to 0.4-3 keV for the final fit of the other model components.

The \textbf{full Galactic} absorption for each HaloSat field is calculated using Planck dust opacity ($\rm \tau_{353}$) maps, as opacity serves as a better tracer of the total absorption column density at the high levels of absorption seen towards the CSB \citep{Planck2014}. The opacity maps are first converted to E(B-V) maps following \citet{Planck2014}. Then the E(B-V) map is converted to $\rm N_H$ following \citet{Zhu2017}. Next, the absorption is weighted by the HaloSat response and the best-fit equivalent $\rm N_H$ value is found by combining the shapes of the weighted absorption curves found over the entire field \citep{LaRocca20202}. This produces a single-valued $\rm N_H$ that more appropriately reflects the range of absorptions over the extended field when compared to a simple average absorption. As the observations are close to the plane of the galaxy, the column densities vary significantly over the entire CSB region, and as such, vary significantly within each HaloSat field. \textbf{This absorption is applied to both the CXB and NEI components.}

The LHB is modeled as a collisionally-ionized diffuse gas {\tt apec} model. The parameters for the LHB {\tt apec} are fixed in XSPEC according to values from \citet{Liu2017} ($kT$ = 0.097 keV), with the normalization calculated from the emission measure. Each emission measure is the average value from the \citet{Liu2017} map for a circular region corresponding to a 5 degree radius (the HaloSat full-response field of view). The CSB region is extrapolated from surrounding areas in \citet{Liu2017} due to infringing X-ray sources making the LHB component difficult to extract in that area. Overall, the LHB is not expected to vary much in any particular direction, so the CSB regions being based on extrapolated emission measures should not be a concern.

The CXB is modeled by an absorbed power-law (XSPEC model {\tt powerlaw}), using \textbf{the} photon index and normalization from \citet{Cappelluti2017}. The normalization is adjusted for the HaloSat field of view. The full Galactic absorption is sufficiently high that the thermal components of the CXB are insignificant.

Another background component stems from the plane of the Galaxy cutting through the middle of the CSB region. It has been documented in multiple \textbf{articles} (e.g. \citet{Iwan1982}, \citet{Valinia1998}, \citet{Rev2006}) that there is significant additional X-ray emission from unresolved sources in the Galactic plane out to as far as 100 degrees from Galactic center. While there is no direct study of the amount of excess emission observed from the galactic ridge within the CSB, the inclusion of a ridge component here would not be unusual (the CSB is roughly 90 degrees from Galactic center). \citet{Kaneda1997} fit the ridge emission with a 2 NEI (XSPEC model {\tt nei}) component model for fields ~30 degrees from Galactic center. Their hard component has model parameters \textbf{of} temperature $kT = 7$ keV, ionization timescale log $n_et$ $\rm cm^{-3}\,s$ $= 10.6$, and abundance $Z = 0.8$. The soft NEI component has model parameters $kT = 0.8$ keV, log $n_et$ $\rm cm^{-3}\,s$ $= 9.1$, and $Z = 0.6$. Both components are included with all parameters fixed to the values from \citet{Kaneda1997}, except for normalization, which is left free for fitting. In most fields, one of the components is significantly observed while the second is only observed as an upper limit. Additionally, different components are detected in different regions of the CSB (although all significantly overlapping fields are consistent). This is not unusual given the large field of view of HaloSat, the differing orientation of each field relative to the plane of the galaxy, and the relatively low scale heights of the NEI components. As an approximation of the true column density, the NEI components were also subjected to full galactic absorption using the same value of $\rm N_H$ as applied to the CXB.

Field HS0016 includes the highly variable X-ray binary Cyg X-3 within its field of view. The MAXI \citep{Matsuoka2009} light curve for Cyg X-3 over the time interval of the HS0016 observations was found to be non-variable. We fit the MAXI data to find the spectral shape of the Cyg X-3 emission at the time of the HaloSat observation. \citet{Hjal2009} mentions a blackbody spectrum combined with a power-law to be a sufficient fit for Cyg X-3, although it is a non-physical simplification. This was found to be a good starting point for a Cyg X-3 fit, with an additional Gaussian component needed to handle the strong iron complex often observed in Cyg X-3 \citep{Kallman2019}. The central value for this Gaussian was found to be similar to the value of a single Gaussian fit for the iron complex in \citet{Hjal2009}. Because Cyg X-3 is a background component rather than a scientific target of interest in this analysis, this model was quite acceptable with a reduced $\rm \chi^2$ of 1.05. The fit parameters found for the MAXI fit (XSPEC models {\tt powerlaw + bbody + gauss}) were as follows: a power-law photon index of 0.0606, a blackbody kT of 1.81 keV, and a Gaussian line energy of 6.46 keV and a Gaussian line width of 0.221 keV. These parameters were then fixed for the Halosat spectra, while the normalizations of the components were allowed to vary while linked to each other in the same ratios as were found in the MAXI fit. This conserves the overall spectral shape of Cyg X-3 in the HaloSat fits. Due to Cyg X-3 being quite a bit further away than the CSB (approximately 3-9 kpc, \citet{Ling2009}), it was also subjected to the same full $\rm N_H$ value as the CXB and NEI components.

\begin{deluxetable*}{ccccccccc}[htb!]
\tablenum{2}
\tablecaption{Model Parameters (0.4-3.0 keV)\label{tab:MP}}
\tablewidth{0pt}
\tablehead{
\colhead{HS ID} & \colhead{l} & \colhead{b} & \colhead{hard NEI norm} & \colhead{soft NEI norm} & \colhead{apec $\rm N_H$} & \colhead{apec kT} & \colhead{apec norm} & \colhead{$\rm chi^2/DoF$} \\
\colhead{}  & \colhead{(deg)}  & \colhead{(deg)}  & \colhead{} & \colhead{} & \colhead{($\rm 10^{22} cm^{-2}$)} & \colhead{(keV)} & \colhead{} &
\colhead{}
}
\startdata
HS0013 & 87.0 & $ -3.3$ & $0.22^{+0.06}_{-0.09}$ & $<0.8$ & $0.55^{+0.06}_{-0.07}$& $0.191^{+0.014}_{-0.008}$ & $77^{+37}_{-29}$ & 165/152 \\
HS0355 & 87.0 & $-4.3$ & $0.13 \pm 0.05$ & $<0.2$ & $0.63^{+0.05}_{-0.06}$ & $0.180^{+0.007}_{-0.006}$ & $140^{+47}_{-28}$ & 198/158 \\
\\
HS0356  & 86.3 & $+5.5$ & $0.17 \pm 0.08$ & $<1.6$ & $0.62^{+0.09}_{-0.08}$ & $0.19^{+0.05}_{-0.01}$ & $58^{+36}_{-26}$ & 259/186 \\
HS0357  & 87.5 & $+6.0$ & $0.07 \pm 0.04$ & $<0.3$ & $0.58^{+0.05}_{-0.06}$ & $0.191^{+0.013}_{-0.008}$ & $52^{+25}_{-18}$ & 220/157 \\
\\
HS0015  & 75.5 & $+9.6$ & $<0.1$ & $0.4^{+0.1}_{-0.3}$ & $0.49 ^{+0.15}_{-0.20}$ & $0.21 \pm 0.03$ & $26^{+46}_{-16}$ & 101/99 \\
HS0358 & 76.5 & $+9.5$ & $<0.03$ & $0.4 \pm 0.1$ & $0.62^{+0.08}_{-0.10}$ & $0.18^{+0.02}_{-0.04}$ & $70^{+28}_{-36}$ & 151/125 \\
\\
HS0016 & 79.7 & $+0.3$ &  $0.37 \pm 0.06$ & $<0.8$ & $0.69 \pm 0.04$ & $0.198^{+0.009}_{-0.007}$ & $82^{+21}_{-18}$ & 357/254 \\
\\
\enddata
\tablecomments{Column 1 is the HaloSat target ID. Columns 2 and 3 are galactic coordinates of the field. Column 4 is the normalization of the 7 KeV NEI component and Column 5 is the normalization of the 0.8 KeV NEI component. Columns 6 through 8 are the CSB {\tt apec} parameters. Column 9 is the $\rm \chi^2$ over degrees of freedom for the fit over the 0.4-3 KeV interval.}
\end{deluxetable*}

\section{Distance Estimates} \label{sec:dis}

A distance must be assumed for further analysis of the properties of the CSB. Previous CSB analyses have been performed with the CSB distance being equivalent to the distance to the Cygnus OB2 association (Cyg OB2). This assumption results in the \textbf{estimated} distance to the CSB changing over time as distance estimates to Cyg OB2 are refined. \citet{Cash1980} estimated a distance of 2 kpc based on their fitted absorption of $\rm 0.56-0.89 \times 10^{22}$ $\rm cm^{-2}$. \citet{Kimura2013} used a distance estimate of 1.7 kpc, predicated upon the CSB being associated with Cyg OB2. \citet{Kimura2013} further justifies linking the distance to OB2 by comparing their fitted absorption of $\rm 0.22-0.33 \times 10^{22}$ $\rm cm^{-2}$ to a set of stars within Cyg OB2 studied in \citet{Yoshida2011}, stating that that paper found a fitted absorption of $\rm 0.2-0.4 \times 10^{22}$ $\rm cm^{-2}$. Unfortunately, the \citet{Yoshida2011} absorptions are fit only for the local circumstellar medium of those OB2 stars, which is combined with a fixed ISM absorption of $\rm 0.83-1.70 \times 10^{22}$ $\rm cm^{-2}$ from \citet{Waldron1998}. Similar high values of extinction of $A_V \sim 5.85-7.07$ ($ N_H \sim 1.44-1.75 \times 10^{22}$ $\rm cm^{-2}$) are found by \citet{Kiminki2015}. As such, the absorption found by \citet{Kimura2013} is not consistent with the Cyg OB2 stellar association. Neither is the absorption found in this paper. Since our measured absorption is inconsistent with the location of the CSB at the distance of Cyg OB2, looking at other OB associations in Cygnus becomes a point of interest.

Cygnus has 9 OB associations within the greater region, but any related OB association is likely to be inside the CSB rather than outside. This limits the associations to Cyg OB1, OB2, OB3, OB4, OB6, OB8, and OB9. \Citet{Uyaniker2001} has an excellent summary of estimated distances as of 2001, but much more has been learned in the last 19 years. Many newer estimates place Cyg OB2 closer, for example, $\rm \sim1.45$ kpc \citep{Hanson2003}, 1.5 kpc \citep{Kharchenko2005}, or $\rm 1.32-1.44$ kpc \citep{Kiminki2015}. \textbf{More recently, the distance to Cyg OB2 has also been studied using Gaia parallax measurements \citep{Gaia2016}. \citet{Lim2019} finds a distance of $\rm 1.6 \pm 0.1$ kpc. \citet{Berlanas2019} finds that Cyg OB2 consists of two populations superimposed at distances of $\rm 1.35^{+0.255}_{-0.220}$ kpc and $\rm 1.76^{+0.396}_{-0.280}$ kpc. Most of the uncertainty in the \citet{Berlanas2019} distance estimates stems from systematic uncertainty in the Gaia parallaxes (see \citet{Stassun2018}).} The OB association summary from \citet{Uyaniker2001} points towards Cyg OB3 being further away than Cyg OB2, and Cyg OB4 being much closer. \citet{Melnik1995} performed a survey of stellar associations and split Cyg OB3 into two associations with distances of 1.82 kpc and 2.31 kpc. Cyg OB3, OB4, and OB6 are also more on the outer edge of the CSB than the other associations. Additional literature on Cyg OB3, OB4, and OB6 is sparse, but the CSB being related to any of them appears unlikely. This leaves Cyg OB1, OB8, and OB9 as \textbf{the} remaining possibilities.

\citet{Melnik1995} found Cyg OB1, OB8, and OB9 to be the same association at a distance of 1.37 kpc. \citet{Kharchenko2005} found a distance for Cyg OB1 of \textbf{1.148 kpc and a distance for Cyg OB9 of 1.139 kpc.} These distance measurements support these associations being interconnected. \citet{Straizys2014} studied stars within the M29 (NGC 6913) cluster, itself found within Cyg OB1, and found a distance for M29 of $\rm 1.54 \pm 0.15$ kpc and a range of extinctions between $A_V = 2.45$ and $A_V = 3.83$ ($ N_H \sim 0.605-0.946 \times 10^{22}$ $\rm cm^{-2}$). This level of absorption found for stars in Cyg OB1 is more consistent with our measurements of the CSB absorption than those found for Cyg OB2.

As a result of these new distances, Cyg OB1 and OB2 would seem to be much closer to each other than previously assumed. There are other observations that provide additional evidence of these associations being close together. S106 is a molecular cloud generally thought to be at $\sim0.6$ kpc \citep{Staude1982} and associated with the Cygnus Rift, although older estimates of much larger distances exist (for example, \citet{Maucherat1975}). \citet{Schneider2007} \textbf{observed} that S106 and kinematically related clouds in the region appear to be interacting with Cyg OB1 and OB2 and thus S106 must be at a distance consistent with those associations. \citet{Giardino2004} found a range of absorption of $\rm \sim 0.8-9.7 \times 10^{22}$ $\rm cm^{-2}$ for potential embedded objects \textbf{in S106}. While those objects are heavily obscured due to their host cloud, the lower end of the absorption range would be consistent with a distance similar to Cyg OB1. The higher absorption seen in Cyg OB2 may be caused by these clouds, with the less absorbed Cyg OB1 positioned more in front of the clouds. This would be consistent with the estimated distances for Cyg OB2 being slightly further away than Cyg OB1. We conclude that the CSB may then be positioned on the near side of Cyg OB1, \textbf{possibly near the foreground population of Cyg OB2 stars from \citet{Berlanas2019},} somewhere between 1.1 and 1.4 kpc.

\section{Results and Discussion} \label{sec:res}

\begin{figure*}[htb!]
\centering
\includegraphics[width=1\textwidth]{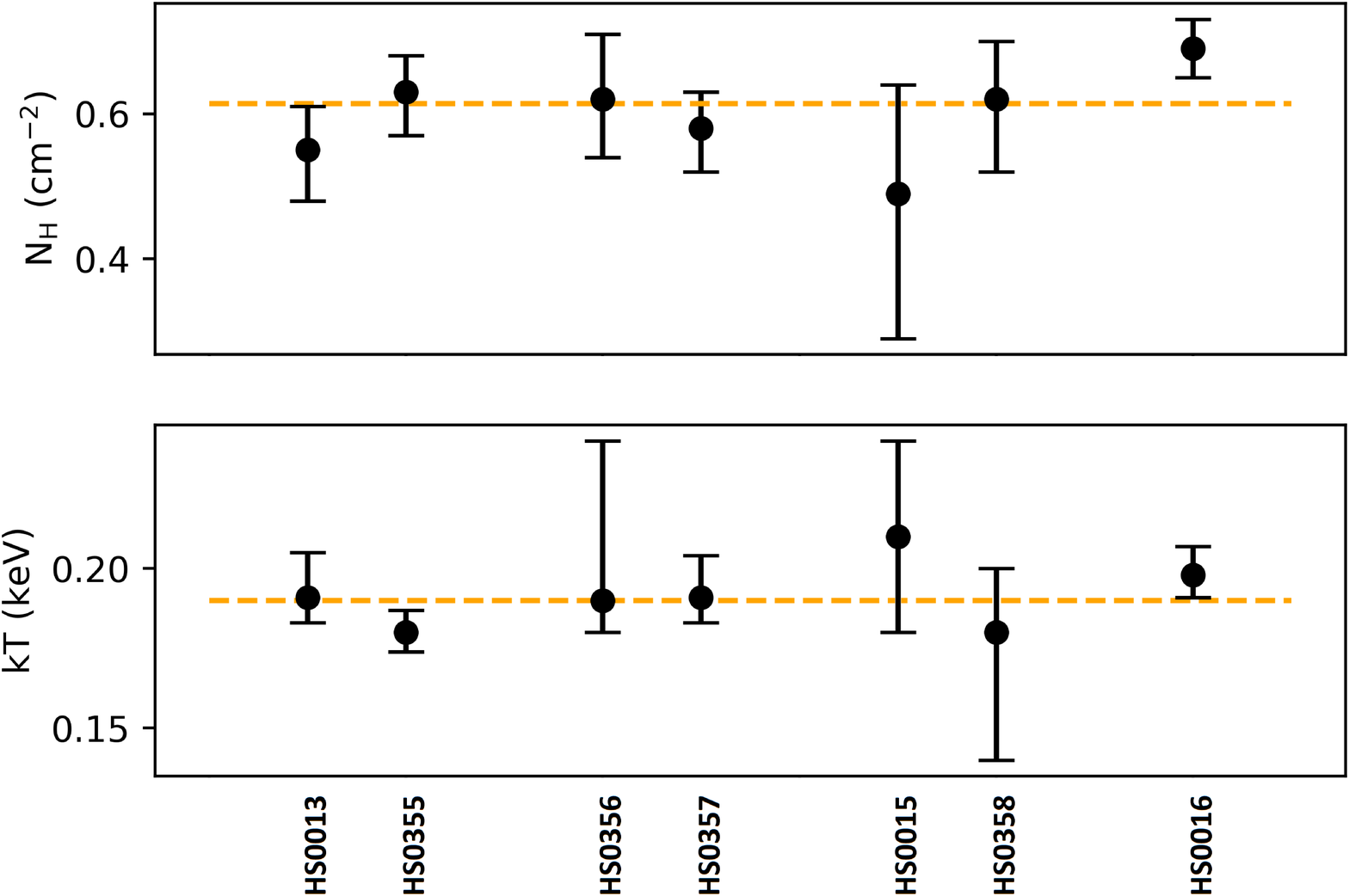}
\caption{Plots of absorption and temperature for the HaloSat CSB fields. Top panel compares the fitted absorption with error for each field to the weighted average absorption (dashed line). Bottom panel compares the fitted temperature with error for each field to the weighted average temperature (dashed line). Overlapping CSB fields are closer to each other than they are to other fields in the plots.}
\end{figure*}

\begin{figure*}[htb!]
\centering
\includegraphics[width=1\textwidth]{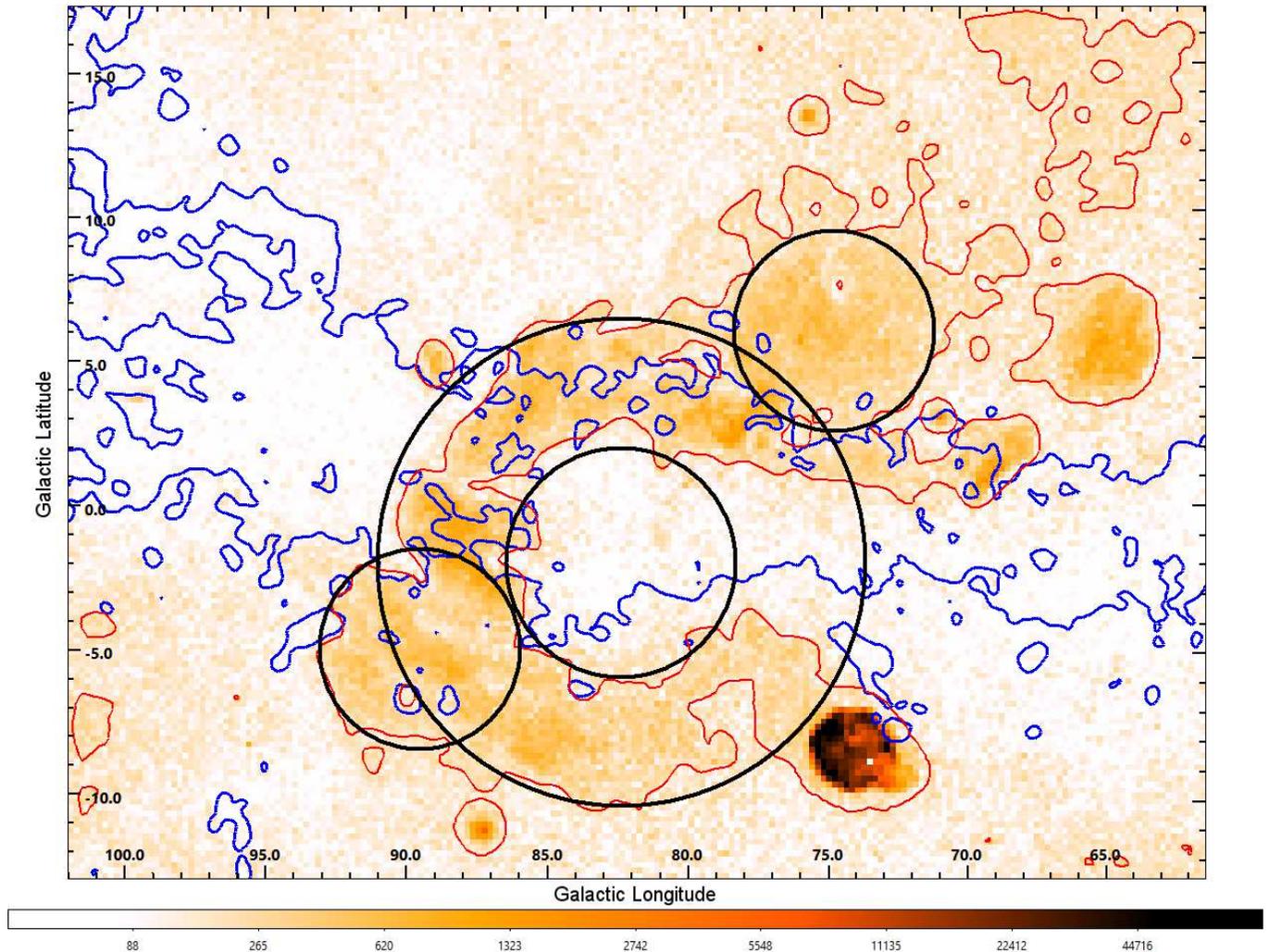}
\caption{ROSAT 3/4 keV (R4+R5) map \citep{Snowden1997} showing the CSB physical model. The CSB emission contour is red. The blue CO contours from \citet{Dame2001} stretch across the middle of the image. The CSB is primarily modeled with a shell, which is the thicker black lines. Two secondary regions to the upper right and lower left are modeled as spheres partially embedded in the shell and are marked with black lines as well. \textbf{The color bar is in units of $\rm 10^{-6}$ ct $\rm s^{-1}$ $\rm arcmin^{-2}$.}}
\end{figure*}

\begin{deluxetable*}{ccccccc}
\tablenum{3}
\tablecaption{Derived Parameters\label{tab:MP}}
\tablewidth{0pt}
\tablehead{
\colhead{HS ID} & Path length & \colhead{Emission measure} & \colhead{Density} & \colhead{Pressure} & \colhead{Thermal Energy} & \colhead{Luminosity}\\
\colhead{} & \colhead{(pc)} & \colhead{($\rm 10^{60} cm^{-3}$)}  & \colhead{($\rm cm^{-3}$)}  & \colhead{($\rm 10^{-11}$ dyne $\rm cm^{-2}$)}  & \colhead{($\rm 10^{51} erg$)} & \colhead{($\rm 10^{37} erg $ $s^{-1}$)}
}
\startdata
HS0013 &$207^{-37}_{+14}$ & $1.8^{-0.6}_{+0.3}$ & $0.067^{+0.007}_{-0.002}$ & $2.05^{+0.21}_{-0.07}$ & $12^{-5}_{+2}$ & $1.9^{-0.6}_{+0.3}$\\
HS0355 &$207^{-37}_{+15}$ & $3.3^{-1.1}_{+0.5}$ & $0.089^{+0.009}_{-0.003}$ & $2.56^{+0.26}_{-0.09}$ & $16^{-6}_{+3}$ & $3.4^{-1.1}_{+0.5}$ \\
\\
HS0356  &$199^{-36}_{+14}$ & $1.4^{-0.4}_{+0.2}$ & $0.078^{+0.008}_{-0.003}$ & $2.37^{+0.25}_{-0.08}$ & $8^{-3}_{+2}$ & $1.4^{-0.5}_{+0.2}$ \\
HS0357  &$189^{-34}_{+14}$ & $1.2^{-0.4}_{+0.2}$ & $0.083^{+0.009}_{-0.003}$ & $2.55^{+0.26}_{-0.09}$ & $7^{-3}_{+1}$ & $1.3^{-0.4}_{+0.2}$ \\
\\
HS0015  &$94^{-16}_{+7}$ & $0.6^{-0.2}_{+0.1}$ & $0.059^{+0.006}_{-0.002}$ & $1.98^{+0.21}_{-0.07}$ & $5^{-2}_{+1}$ & $0.6^{-0.2}_{+0.1}$ \\
HS0358  &$95^{-17}_{+7}$ & $1.6^{-0.5}_{+0.2}$ & $0.096^{+0.010}_{-0.003}$ & $2.77^{+0.29}_{-0.09}$ & $7^{-3}_{+1}$ & $1.7^{-0.6}_{+0.3}$ \\
\\
HS0016  &$175^{-31}_{+12}$ & $1.9^{-0.6}_{+0.3}$ & $0.082^{+0.009}_{-0.003}$ & $2.61^{+0.25}_{-0.10}$ & $11^{-4}_{+2}$ & $2.0^{-0.6}_{+0.3}$ \\
\\
Full CSB & ... & ... & $0.079^{+0.008}_{-0.003}$ & $2.41^{+0.25}_{-0.08}$ & $39^{-17}_{+9}$ & ... \\
\\
\enddata
\tablecomments{This table includes parameters derived from the best fit model parameters in XSPEC for each field, as well as estimated global parameters for the entire CSB, assumed here as a cohesive object, using average values found for the HaloSat observations. The values are calculated assuming a distance of 1.4 kpc, while the listed error is calculated by using different distance estimates. The upper error value is the difference between 1.15 kpc and 1.4 kpc distances, while the bottom error bar is the difference between 1.4 kpc and 1.5 kpc. The difference between a filled spherical versus shell like model is roughly an additional 10\% error.}
\end{deluxetable*}

The fitted model parameters for absorption and temperature in the different HaloSat fields (Table 2) are consistent with each other. Taking the weighted average for absorption returns a value of $\rm 0.614 \times 10^{22}$ $\rm cm^{-2}$ for the data having a $\rm \chi^2$/dof deviation from the average of 5.6/6. The weighted average for temperature is 0.190 keV with a $\rm \chi^2$/dof deviation for the data from the average of 3.9/6. Figure 4 is a plot of the fitted absorptions and temperatures, and includes a comparison to the weighted averages.

While the best-fit column densities for HaloSat do not match those from \citet{Kimura2013}, the values remaining consistent across the CSB regions still reinforces the conclusion that the CSB is a single structure. The \citet{Kimura2013} model fits are simultaneous fits of ROSAT and MAXI spectra, and the resulting model fits for the MAXI data exhibit excess emission at the lowest energies (0.7-0.8 kev). The HaloSat data goes down to 0.4 \textbf{keV, exhibits no such issue with excess emission (see Figures 2 and 3),} and avoids potential sources of error stemming from simultaneously fitting two separate observatory's data. Absorption has a significant effect below 0.5 keV, so the energy range of HaloSat is sufficient in this regard. Also note that while the calculated column densities from HaloSat exceed \citet{Kimura2013} by a fair margin, they are within the range of column densities ($\rm 0.15-1.2 \times 10^{22}$ $\rm N_{H}$) found by \citet{Uyaniker2001}, and are consistent with the absorption of $\rm 0.56-0.89 \times 10^{22}$ $\rm cm^{-2}$ found by \citet{Cash1980}.

Using the best-fit spectral parameters in Table 2, several additional physical characteristics of the CSB can be calculated. First some initial assumptions about a physical model of the CSB must be made. As seen in Figure 5, the CO contours \citep{Dame2001} in this region trace gaps in emission (the CO contours also trace the dust opacity from \citet{Planck2014}) and provide evidence that the southern arc of the CSB is not bound by absorption, rather it is an expression of the actual structure of the CSB. This also appears to be true of the upper edge of the northern arc as well. Due to the strength of the emission dropping off in the CSB interior, the central volume of the CSB is assumed to form a shell and the thickness of the shell is based on the thickness of this southern arc, matched with the outer boundary of the northern arc. Gaps in the shell are assumed to be absorbed by the intervening dust, which the CO contours trace. Two significant regions of emission remain outside the shell, and these have been assumed to be spherical as they exhibit no obvious sign of being additional shell-like remnants. These two regions may be separate supernova remnants in the region, or might be low-density voids where the CSB expansion has propagated faster. \textbf{The CSB shell model is centered at galactic coordinates $l=82.5^{\circ}$ and $b=-2.0^{\circ}$, with the two spherical regions at $l=75.0^{\circ}$, $b=+6.0^{\circ}$ and $l=89.5^{\circ}$, $b=-5.0^{\circ}$. The outer radius of the shell is 8.5 degrees and the inner radius is 4.0 degrees. The secondary spheres each have a radius of 3.5 degrees.}

In order to calculate the physical parameters of interest, an estimate of path length through the observed sections of the CSB is needed. This is done by generating a grid of path lengths across the CSB model and removing values that lie outside the CSB contour under the aforementioned assumption that the emission from those parts are heavily absorbed. The two additional spherical components are treated as being embedded in the shell, so for any given point the greater path length of the two shapes is used. An average path length is found for each observed CSB field. This average is found over the entire field, with the path lengths reduced proportionally for any portions of the model that lie in the region of the field with a reduced response. This procedure is repeated for 3 separate possible distances of 1.15 kpc, 1.4 kpc, and 1.5 kpc based upon the distance estimates for Cyg OB1 and OB2 discussed in section 4. 

Table 3 includes the following parameters for each field: path length, emission measure, density, pressure, thermal energy, and luminosity. The error ranges of these values are based upon the aforementioned range of distances. The luminosity for each field is calculated by finding the flux in XSPEC by using a dummy response and the {\tt apec} model for the field, unabsorbed, over an energy range of 0.1-20 keV. This is repeated for the three previously mentioned distances. The table values would increase by up to $\sim $10\% if the emitting region were a filled sphere rather than a shell, depending on how much of the shell is in the field of view.

Despite the significant differences in what part of the CSB model is captured by each field, the derived parameters are all rather similar, lending support to both the CSB being a hypernova remnant as well as support for the shell remnant model. The similar values for density can be averaged and combined with the weighted average temperature to calculate global values for the entire CSB. These values are also included in Table 3. Taking the thermal energy for each field and dividing it by the luminosity of the field returns an estimate of cooling time (calculated using the 1.4 kpc distance estimate). Averaging all the fields, the CSB has an estimated cooling time of 18 million years. Note, however, that this is functionally an upper bound to the cooling time that does not take into account energy losses due to expansion and radiation at other wavelengths. This estimate corresponds to an energy loss of $\rm \sim 4 \times 10^{52}$ $\rm erg$ over that time, based on an estimated total luminosity of the CSB of $\rm \sim 7 \times 10^{37}$ $\rm erg$ $\rm s^{-1}$. This is an initial energy of $\rm \sim 8 \times 10^{52}$ $\rm erg$, similar to the $\rm 2-5 \times 10^{52}$ erg found for hypernova SN1998bw by \citet{Iwamoto1998}.

\textbf{If the CSB is a hypernova remnant, then the question remains of where the progenitor star came from. The relative positions of the CSB model and the Cyg OB associations can be seen in Figure 6. BD+43 3654, a 70 solar mass runaway star ejected from Cyg OB2, has proper motion of 5.7 mas $\rm yr^{-1}$ and an estimated age of 1.6 Myr \citep{Comeron2007}. This corresponds to a total distance traveled of $\rm \sim 2.5$ degrees on the sky, which is similar to the distance between the CSB center and Cyg OB2. Depending on the exact distance to the CSB, the closer Cyg OB2 population from \citet{Berlanas2019} could be the source of an ejected CSB progenitor star.}

\begin{figure*}[htb!]
\centering
\includegraphics[width=1\textwidth]{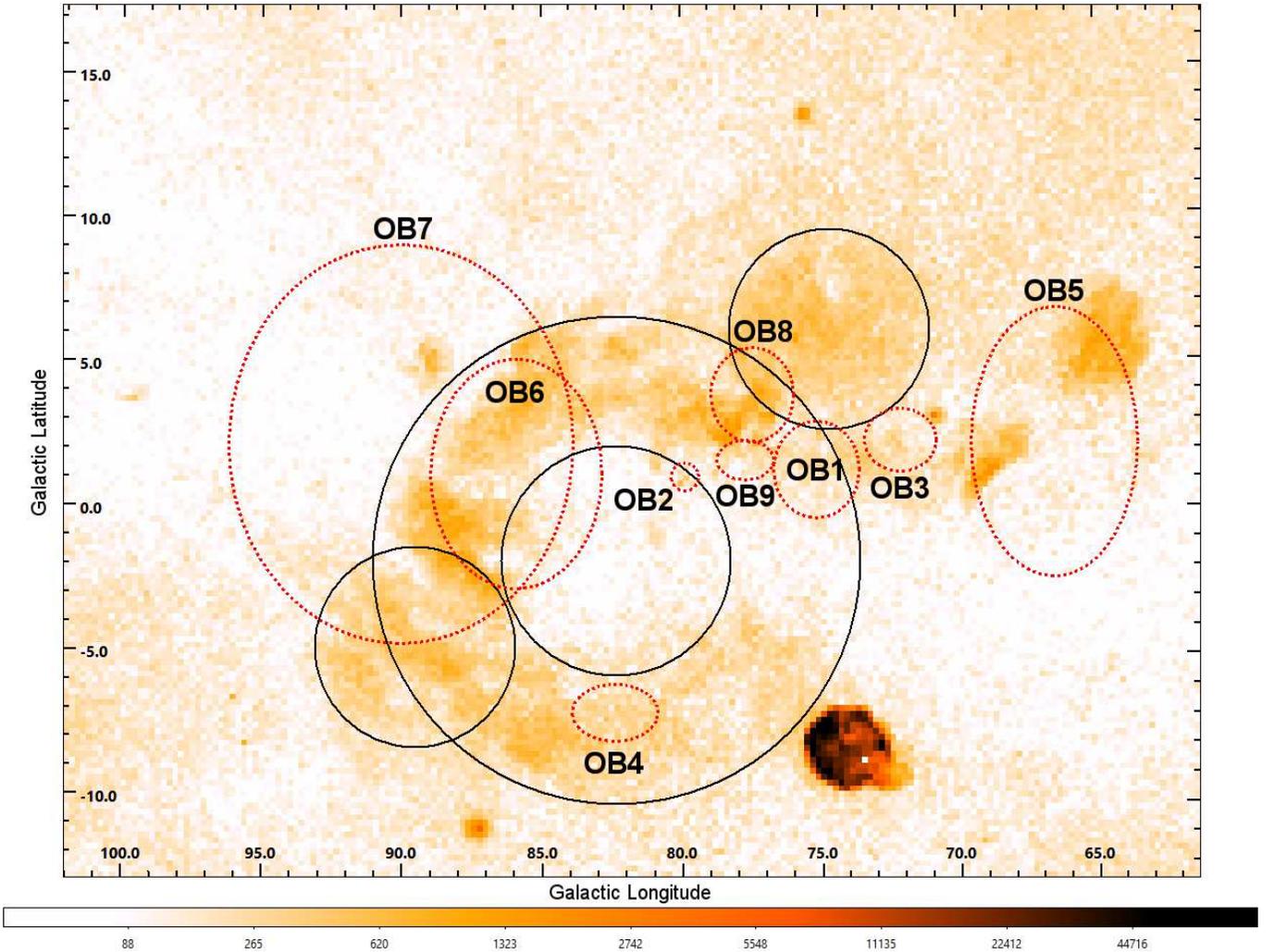}
\caption{\textbf{ROSAT 3/4 keV (R4+R5) map \citep{Snowden1997} showing the CSB physical model in black and the Cyg OB associations in red dashed lines. The coordinates and sizes of the OB associations are taken from Table 1 in \citet{Uyaniker2001}, with Cyg OB4's latitude corrected to the original value from \citet{Humphreys1978}. The diameter used for Cyg OB2 is from \citet{Knodlseder2000}. The color bar is in units of $\rm 10^{-6}$ ct $\rm s^{-1}$ $\rm arcmin^{-2}$.}}
\end{figure*}

The study of the Cygnus OB associations (OB1, OB3, OB7, and OB9) from \citet{Comeron1998} finds association ages of only around 5-10 million years, and their proper motion measurements show that stars in the associations are moving \textbf{away from the center of the CSB}, interpreting this as a sign that the expansion of the CSB triggered the formation of the Cygnus association stars. This relationship is supported by this paper's absorption measurements, meaning that the relation between the CSB and Cyg OB1 may not be that Cyg OB1 gave rise to the CSB progenitor star, rather that the CSB triggered the formation of Cyg OB1. \textbf{Cyg OB2 also exhibits proper motion away from the center of the CSB \citep{Lim2019}, so triggered star formation caused by the CSB expansion may also be happening in Cyg OB2.} 

\textbf{\citet{Comeron2016} finds that Cyg OB2 exhibits a older secondary population of stars, with an age of 20 Myr, based on observations of red supergiants in the region. This would have interesting implications for Cyg OB2 as a potential source of a progenitor star, especially in combination with the results from \citet{Berlanas2019}. However, the majority of the Gaia DR2 parallaxes for the \citet{Comeron2016} red supergiant sample would be considered foreground/background contaminants when added to the \citet{Berlanas2019} sample. Only RAFGL 2600 and IRAS 20341+4047 have parallaxes that are consistent with Cyg OB2, with those stars respectively falling in the 1.35 kpc and 1.76 kpc populations.} 

\begin{deluxetable}{cc}
\tablenum{4}
\tablecaption{Parallaxes for Candidate Cyg OB2 Stars \label{tab:MP}}
\tablewidth{0pt}
\tablehead{
\colhead{Star} & Parallax \\
\colhead{} & \colhead{(mas)} 
}
\startdata
IRAS 20315+4026 & $\rm -0.256 \pm 0.114$ \\
IRAS 20263+4030 & $\rm 1.079 \pm 0.122$  \\
RAFGL 2605 & $\rm 1.022 \pm 0.134$ \\
IRAS 20290+4037 & $\rm 1.175 \pm 0.095$ \\
RAFGL 2600 & $\rm 0.753 \pm 0.100$ \\
IRAS 20249+4046 & $\rm 1.033 \pm 0.129$ \\
IRAS 20341+4047 & $\rm 0.570 \pm 0.135$ \\
\enddata
\tablecomments{Gaia DR2 parallaxes \citep{Gaia2018} for the 7 stars studied in \citet{Comeron2016}. Gaia systematic error is not included in these measurements.}
\end{deluxetable}

The total thermal energy of the CSB at a distance of 1.4 kpc is estimated to be $\rm 3.9\times 10^{52}$ erg. This is similar to the total thermal energy found by \citet{Kimura2013} of $\rm 9 \times 10^{51}$ erg, although they assumed a larger distance for the CSB. The reason that the total thermal energy ends up being similar is due to a combination of a higher calculated density (0.08 $\rm cm^{-3}$ versus the \citet{Kimura2013} value of 0.02 $\rm cm^{-3}$) and a significantly different geometry of emitting region for the CSB resulting in a larger volume ($\rm \sim 1 \times 10^{63}$ $\rm cm^{3}$ or $\rm \sim 38,000,000$ $\rm pc^{3}$). \citet{Kimura2013} uses a C-shaped toroidal model in the plane of the observation, with a volume of $\rm ~4 \times 10^{62}$ $\rm cm^{3}$. The \citet{Kimura2013} model is motivated only to cover the observed emission, rather than attempting to describe the full highly-absorbed structure. 

The consistent parameters found for the CSB emission, both fit and derived, imply that it stems from a singular origin. If the interior volume of the CSB was blown out by varying degrees of stellar winds and/or extended supernova activity, then there is no particular expectation that these different sides of the CSB would feature similar parameter values, given their differing points of origin physically and chronologically. This makes the hypernova interpretation of the CSB origin more favorable, as the edges of the CSB would thus be expected to have similar history in terms of starting energy and cooling times.

\section*{Acknowledgements}
This research was supported by NASA grant No. NNX15AU57G. This research has made use of MAXI data provided by RIKEN, JAXA and the MAXI team. This work has made use of data from the European Space Agency (ESA) mission {\it Gaia} (\url{https://www.cosmos.esa.int/gaia}), processed by the {\it Gaia} Data Processing and Analysis Consortium (DPAC, \url{https://www.cosmos.esa.int/web/gaia/dpac/consortium}). Funding for the DPAC has been provided by national institutions, in particular the institutions participating in the {\it Gaia} Multilateral Agreement.

\software{DS9 \citep{Joye2003}, matplotlib \citep{Hunter2007}, NumPy \citep{Harris2020}, XSPEC (v12.10.1f; Arnaud 1996)}


{}


\end{document}